\documentclass[3p,twocolumn,preprint, sort&compress]{elsarticle}

\journal{Physics Letters B}

\usepackage{amsmath,amsfonts,amssymb}  
\usepackage{booktabs}
\usepackage{graphicx} 
\usepackage[colorlinks=true,urlcolor=black,citecolor=blue,linkcolor=black] {hyperref}

\usepackage{xcolor}

\newcommand{\be}{\begin{equation}}
\newcommand{\ee}{\end{equation}}
\newcommand{\br}{\begin{eqnarray}}
\newcommand{\bea}{\begin{eqnarray}}
\newcommand{\eea}{\end{eqnarray}}
\newcommand{\er}{\end{eqnarray}}
\newcommand{\ba}{\begin{array}}
\newcommand{\ea}{\end{array}}
\newcommand{\bi}{\begin{itemize}}
\newcommand{\ei}{\end{itemize}}
\newcommand{\bn}{\begin{enumerate}}
\newcommand{\en}{\end{enumerate}}
\newcommand{\bc}{\begin{center}}
\newcommand{\ec}{\end{center}}

\newcommand{\abs}[1]{\lvert #1 \rvert}

\begin{document}
\begin{frontmatter}

\title{Dark Matter-Induced Multi-Phase Dynamical Symmetry Breaking}

\author[a]{Kristjan Kannike\corref{mycorrespondingauthor}}
\cortext[mycorrespondingauthor]{Corresponding author}
\ead{kannike@cern.ch}
\author[a]{Niko Koivunen}
\author[a]{Aleksei Kubarski}
\author[a]{Luca Marzola}
\author[a]{Martti Raidal}
\author[b]{Alessandro Strumia}
\author[a]{Venno Vipp}
\address[a]{NICPB, R\"avala 10, 10143 Tallinn, Estonia}
\address[b]{Dipartimento di Fisica, Universit\`a di Pisa, Italia}

\begin{abstract} 
We consider the classically scale invariant Higgs-dilaton model of dynamical symmetry breaking 
extended with an extra scalar field that plays the role of dark matter. 
The Higgs boson is light near a critical boundary between different symmetry breaking phases, where
quantum corrections beyond the usual Gildener-Weinberg approximation become relevant.
This implies a tighter connection between dark matter and Higgs phenomenology.
The model has only three free parameters, 
yet it allows for the observed relic abundance of dark matter while respecting all constraints.
The direct detection cross section mediated by the Higgs boson is determined by the dark matter mass alone
and is testable at future experiments.
\end{abstract}


\end{frontmatter}

\section{Introduction}
\label{sec:intro}

The Coleman-Weinberg mechanism for dynamical symmetry breaking predicts a scalar lighter than
the scale of dimensional transmutation: the pseudo-Goldstone boson of broken scale invariance, dubbed the `dilaton'.
The original Coleman-Weinberg proposal identified dimensional transmutation with breaking of the electroweak symmetry and, thereby, the dilaton with the Higgs boson~\cite{Coleman:1973jx}. This possibility is now excluded, so models allowed by current data involve an extra scalar field. In view of the absence of new physics around the electroweak scale, recent analyses have focused on  a limit where also the Higgs boson is lighter than the scale of dimensional transmutation.

As recently shown in~\cite{Kannike:2021iyh}, the lightness of the Higgs boson signals proximity to a critical boundary where two different phases of dynamical symmetry breaking overlap and extra quantum effects, neglected by the usual Coleman-Gildener-Weinberg approximation~\cite{Gildener:1976ih}, become relevant. In detail, near the multi-phase criticality point the Higgs boson mass is suppressed by scalar $\beta$-functions similarly to the dilaton mass, allowing one to create a large hierarchy between masses of different scalars. At the same time, the dilaton, which can be heavier or lighter than the Higgs boson, mixes only weakly with the latter. The proposed multi-phase criticality scenario is consistent with the fact that no new physics has been found around the electroweak scale despite the existence of several light and heavy scalars predicted by the scenario~\cite{Huitu:2022fcw}.



In the present work we extend the minimal multi-phase criticality scenario of~\cite{Kannike:2021iyh,Huitu:2022fcw} by adding a further scalar field as the dark matter (DM) candidate. The scalar sector of the model we consider thus contains the Standard Model (SM) Higgs doublet and two additional fields, 
neutral under the SM gauge group.\footnote{Singlet scalar DM in such a model, without dynamical symmetry breaking, was studied in~\cite{Abada:2011qb,Abada:2012hf,Abada:2013pca}.} 
This is an explicit realization of the generic scenario of~\cite{Kannike:2021iyh,Huitu:2022fcw} in which the DM is responsible for dynamical symmetry breaking in the vicinity of the multi-phase critical point. The same model was previously investigated with the usual Gildener-Weinberg approach in~\cite{Ishiwata:2011aa}. We go beyond this approximation and show that around the multi-phase critical point new effects occur.

The phenomenology of our model is predicted in terms of three free parameters: the DM mass, the dilaton mass and a parameter that connects the usual Gildener-Weinberg limit with the solutions obtained in the multi-phase criticality regime. As we show below, this parameter uncovers a deeper connection between DM physics and electroweak symmetry breaking. To assess the phenomenology of the model, we account for the constraints on the DM relic density, direct detection cross section and the latest collider bounds. We find that there exist several parameter regions, both for small and large dilaton masses, for which the model avoids all the existing bounds and predicts the observed DM relic density. Most interestingly, the DM direct detection cross section, although mediated by the Higgs boson, is determined only by the DM mass. This property reveals the connection between the Higgs dynamics and the large scalar DM mass. We conclude that the model is potentially testable at future direct detection and collider experiments.




\section{Model}
\label{sec:model}

We extend the SM by adding two singlet scalar fields denoted by $s$ and $s'$. The scale-invariant scalar potential, symmetric under a $\mathbb{Z}_2\otimes\mathbb{Z}'_2$ symmetry
acting on the new states as $s\to - s$ and as $s'\to - s'$, is 
\begin{equation}
\begin{split}
  \label{eq:potential}
  V &= \lambda_{H} |H|^{4} + \frac{ \lambda_{S}}{4}  s^{4} + 
   \frac{\lambda_{S'} }{4} s^{\prime 4} + \frac{ \lambda_{HS}}{2} |H|^2  s^{2} 
   \\
  &+ \frac{ \lambda_{HS'} }{2} |H|^2 s^{\prime 2}
  + \frac{ \lambda_{SS'}}{4} s^{2} s^{\prime 2}.
\end{split}
\end{equation}
The scalar $s$ acquires a vacuum expectation value (VEV) $w$
through the Coleman-Weinberg mechanism which spontaneously breaks its $\mathbb{Z}_2$ and the classical scale symmetry. 
Thereby $s$ mixes with the Higgs and decays.
The DM candidate, $s'$, does not acquire a VEV and is stabilized by the unbroken $\mathbb{Z}'_2$ symmetry.

Given the requirement of vanishing VEV for $s'$, realised for $\lambda_{S'}, \lambda_{HS'}, \lambda_{SS'} > 0$ (for necessary and sufficient conditions see~\cite{Kannike:2012pe}), the potential in Eq.~\eqref{eq:potential} admits the following phases:

\begin{enumerate}
\item[$s$)] $\lambda_S =0$ and $\lambda_{H}, \lambda_{HS}>0$ for $w\neq 0$ and $v=0$;

\item[$h$)] $\lambda_H =0$ and $\lambda_{S}, \lambda_{HS}>0$ for $v\neq0$ and $w=0$;

\item[$sh$)] $\lambda_{HS} =-2\sqrt{\lambda_H \lambda_S}< 0$
and $\lambda_{H}, \lambda_{S}>0$ for $w,v\neq 0$,
\end{enumerate}
where $v$ denotes the vacuum expectation value of the Higgs boson.  
The $sh$ phase is compatible with present constraints and yields the usual Gildener-Weinberg construction if the renormalization group (RG) running of $\lambda_{HS}$ can be neglected. In such a case the classical potential is flat along the direction 
\begin{equation} h/s = \sqrt{-\lambda_{HS}/2\lambda_H},
\end{equation}
and only quantum corrections along this direction are included in the Gildener-Weinberg approximation.
The coupling $\lambda_{HS'}$ between DM and the Higgs boson is important for DM phenomenology, in particular for direct detection, but in this regime it does not enter the dynamics of symmetry breaking. The DM phenomenology of this model has been studied in~\cite{Ishiwata:2011aa} and, in view of the present Higgs physics bounds, small values of $\lambda_{HS}$ are to be considered as the parameter regulates the mixing between the Higgs boson and the dilaton.

A deeper connection between DM and symmetry breaking
arises if $\lambda_{HS}$ is so small that quantum corrections to it
become relevant, invalidating the usual Gildener-Weinberg approximation. Such a regime is unavoidably realized at the multi-phase critical limit
where the two phases, $s$ and $sh$, merge:
\begin{equation}
  \lambda_{S}(\bar\mu) = 0, \qquad \lambda_{HS}(\bar\mu) \approx 0
  \label{eq:multi:critical:lambdas}.
\end{equation}
The second equation holds only approximately at the symmetry breaking scale, since the Higgs boson is massless
at the critical boundary ($\lambda_{HS}(\bar\mu)=0$). 
This multi-phase criticality scenario~\cite{Kannike:2021iyh} takes into account the quantum corrections to $\lambda_{HS}$ that push the true minimum of the potential away from the direction indicated by the Gildener-Weinberg approach \cite{Alexander-Nunneley:2010tyr,Kannike:2020ppf}. Whereas these contributions are generally negligible, in the multi-phase criticality limit they affect electroweak symmetry breaking and Higgs boson mass generation. 

\smallskip

The lightness of the Higgs boson around the multi-phase critical point is phenomenologically motivated, given the non-observation of new physics around the weak scale. Furthermore, it shows that in order to compute the correct scalar mass spectrum and mixing, one needs to go beyond the Gildener-Weinberg approximation and include the loop corrections to $\lambda_{HS}$. In geometric terms the situation can be understood as follows: the value of $\lambda_{HS}$ determines the Gildener-Weinberg approximately flat direction; so the relevance of its renormalization group (RG) running means that the potential is approximately flat along a curved region in field space, rather than along a straight segment. The displacement of the minimum is one consequence of this.

Due to the multi-phase criticality condition in Eq.~\eqref{eq:multi:critical:lambdas}, the contribution of $\lambda_{HS}$ to its own running is negligible and an additional field is needed to drive the running. In the present model, it is the DM candidate $s'$ that provides the largest contributions to the
$\beta$-functions of the parameters that determine which symmetry breaking phase is realized:\footnote{The full one-loop expressions of the $\beta$-functions are given in~\ref{sec:RGEs}. The terms proportional to $\lambda_{HS}$, neglected in the approximations, contribute up to $\mathcal{O}(10\%)$ in the considered parameter ranges.}
\begin{equation}\beta_{\lambda_{HS}}\simeq \frac12 \lambda_{SS'} \lambda_{HS'},\qquad
  \beta_{\lambda_S} \simeq \frac14 \lambda_{SS'}^2.
  \label{eq:beta:at:s0}
\end{equation}
Therefore, in the multi-phase criticality regime, the coupling $\lambda_{HS'}$ between DM and the Higgs  (crucial for DM direct direction) also affects the symmetry breaking dynamics.

\medskip

\begin{figure}[tb]
\begin{center}
  \includegraphics{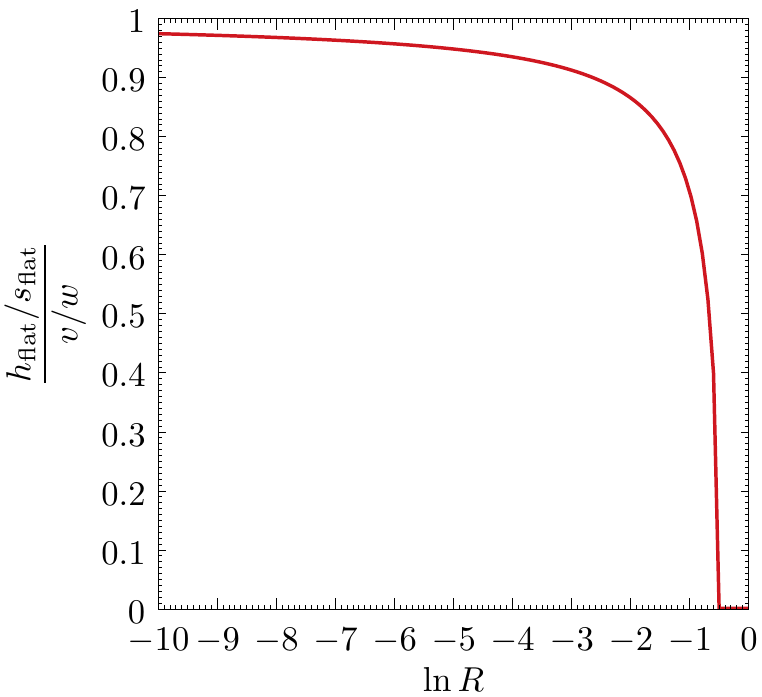}
\caption{\em The ratio of the flat direction angle to the actual minimum direction angle as a function of the $\ln R$ parameter.}
\label{fig:flat:vev}
\end{center}
\end{figure}

As we will show below, experimental constraints select a region of the parameter space where the dilaton VEV is well above the electroweak scale and the DM mass is much larger than those of the remaining scalar states. Therefore, near its minimum, the potential at one-loop is controlled by the DM field-dependent mass 
\be m_{s'}^2(s,h) = \frac{1}{2} (\lambda_{SS'} s^2 + \lambda_{HS'} h^2)
\ee
that dominantly depends on the $s$ field value. The improved effective potential is thus well approximated by its tree-level form augmented with
$s$-dependent running couplings $\lambda_S$ and $\lambda_{HS}$~\cite{Kannike:2021iyh}. Importantly, since the DM mass $m_{s'}$ is dynamically generated and receives no field-independent inputs, the contributions of $s'$ to the $\beta$-functions of Eq.~\eqref{eq:beta:at:s0} must be retained even at field values below $m_{s'}$, violating na\"{i}ve decoupling expectations. Consequently, we have  
\begin{equation}
\lambda_S^{\rm eff}(s) = \frac{{\beta}_{\lambda_S }}{(4\pi)^2}   \ln \frac{s^2}{s^2_S },
\label{eq:run:lambda}
\end{equation}
\begin{equation}\label{eq:defR}
\lambda_{HS}^{\rm eff}(s) = \frac{{\beta}_{\lambda_{HS} }}{(4\pi)^2} \ln \frac{R s^2 }{e^{-1/2}s^2_{S}},
\end{equation}
where $s_S$ is the flat-direction scale and the parameter $\abs{R} \sim 1$ across the region near the multi-phase criticality boundary (the $e^{-1/2}$ factor is included to simplify the subsequent equations).\footnote{The parameter $R$ takes into account both the tree-level values of the quartic couplings and the finite shifts due to the one-loop part of the effective potential. In place of $R$, we could write the above equation as
\begin{equation*}
\lambda_{HS}^{\rm eff}(s) = 
\lambda_{HS}^0+\frac{{\beta}_{\lambda_{HS} }}{(4\pi)^2} \ln \frac{s^2 }{s^2_{S} e^{-1/2}}.
\end{equation*}
to make transparent that the usual Gildener-Weinberg is recovered in the limit where
$\lambda_{HS}^0\equiv\lambda_{HS}^{\rm eff}(w)$
is large enough that the ${\beta}_{\lambda_{HS}}$ running can be neglected. The parametrisation in terms of $R$, however, is more suited to the purposes of the present analysis.} 
Here $w$ and $v$ are the VEVs of $s$ and $h$, respectively, given by
\begin{equation}
w \equiv s_S \, e^{-1/4},\qquad
v = \frac{w}{4 \pi} \sqrt{-\frac{\beta_{\lambda_{HS}} \ln R}{2\lambda_H}} .
\label{eq:vevs}
\end{equation}
The minimum is displaced from the Gildener-Weinberg flat direction by an amount that becomes relevant in the multi-phase criticality limit
\begin{equation}
  \frac{ h_{\rm flat}}{ s_{\rm flat}} {\Bigg /} \frac{v}{w} = \sqrt{1 + \frac{1}{2 \ln R}},
  \label{eq:flat:over:min}
\end{equation}
as illustrated in Fig.~\ref{fig:flat:vev}. Notice that this ratio also equals  $\sqrt{\lambda_{HS}^{\rm eff}(s_{\rm flat})/\lambda_{HS}^{\rm eff}(w)}$, confirming that the Gildener-Weinberg regime is obtained in the limit where the running of the Higgs portal $\lambda_{HS}$ can be neglected. 

\subsection{A phenomenological parametrization}

The potential in Eq.~\eqref{eq:potential} contains six parameters. 
Two are set by the measurements of the Higgs  mass, $m_h \approx \sqrt{2 \lambda_{H}} v \approx 125.2$ GeV, 
and vacuum expectation value, $v=246.2$ GeV. Since the parameter $\lambda_{S'}$ only determines the self-interactions of DM, we are effectively left with three free parameters, which we can trade for the dilaton mass $m_{s}$, the DM mass $m_{s'}$ and  $\ln R$. When $\ln R$ is large and negative, it describes a constant value of $\lambda_{HS}$ that neglects its RG running, see Eq.~\eqref{eq:defR}.

We next express the DM couplings $ \lambda_{SS'}$ and $ \lambda_{HS'} $ in terms of our free parameters. An exact parametrization is given in~\ref{sec:exact}. The mass matrix elements for the $h$ and $s$ fields are given in Eqs.~\eqref{eq:m:hh}, \eqref{eq:m:ss} and \eqref{eq:m:h:s}. Above the lower bound on $m_{s'}$ given by Eq.~\eqref{eq:msp:lower:bound}, we can neglect the $v^{2}$ terms in the mass matrix and approximate the eigenvalues as
\begin{align}
  m^2_{h} & \simeq - \frac{\beta_{\lambda_{HS}}}{(4 \pi)^{2}} w^2 \ln R, 
   \label{eq:m:sq:h}
  \\
  m^2_{s} & \simeq 2  \frac{\beta_{\lambda_S}}{(4 \pi)^{2}}  w^2,
   \label{eq:m:sq:s}
  \\
  m^{2}_{s'} &\simeq \frac{1}{2} \lambda_{SS'} w^{2},
  \label{eq:m:sq:sp}
\end{align}
while the mixing angle is given by
\begin{equation}
\theta\simeq \frac{m_{hs}^2}{m_{s}^2-m_{h}^2}
=\frac{\beta_{\lambda_{HS}} (1 + \ln R)}{2 \beta_{\lambda_{S}} + \beta_{\lambda_{HS}} \ln R} \frac{v}{w},
\label{eq:theta}
\end{equation}
where $m_{hs}^2$ is the mixing mass term among $h$ and $s$.

With the approximations in Eqs.~\eqref{eq:m:sq:h}, \eqref{eq:m:sq:s} and \eqref{eq:m:sq:sp}, the expressions of the quartic couplings in Eqs.~\eqref{eq:DM:dilaton:portal:compl} and \eqref{eq:DM:Higgs:portal:compl} simplify as
\begin{align}
  \lambda_{SS'} &\approx \frac{(4 \pi)^{2} m_{s}^{2}}{m_{s'}^{2}},
  \label{eq:DM:dilaton:portal}
\\
  \lambda_{HS'} &\approx -\frac{(4 \pi)^{2} m_{h}^{2}}{m_{s'}^{2} \ln R}.
  \label{eq:DM:Higgs:portal}
\end{align}
Notice that while $m_{s'}^2 \simeq \lambda_{SS'} w^2 / 2$,
the coupling $\lambda_{SS'}$ diminishes with larger values of $m_{s'}$ due to the dependence of the dilaton VEV $w$ on $m_{s'}$:
\begin{equation}
  w \simeq \frac{\sqrt{2}{m_{s'}^{2}}}{{4 \pi m_{s}}}.
  \label{eq:w}
\end{equation}
Finally, the mixing angle between $h$ and $s$ in Eq.~\eqref{eq:theta} can be approximated as
\begin{align}
\theta\simeq &\frac{2 \sqrt{2} \pi m_{s} m_{h}^{2} v (1 + \ln R)}{(m_{h}^{2} - m_{s}^{2}) m_{s'}^{2} \ln R}.
\end{align}
We next assess the phenomenology of the model, considering values of $-\ln R$ within a range that spans from unity (multi-phase criticality regime) to the large values that recover the Gildener-Weinberg limit. After that, focusing on the multi-phase criticality, we assess the impact of DM phenomenology on the parameter space of the model.

\begin{figure*}[ptb]
\begin{center}
  \includegraphics{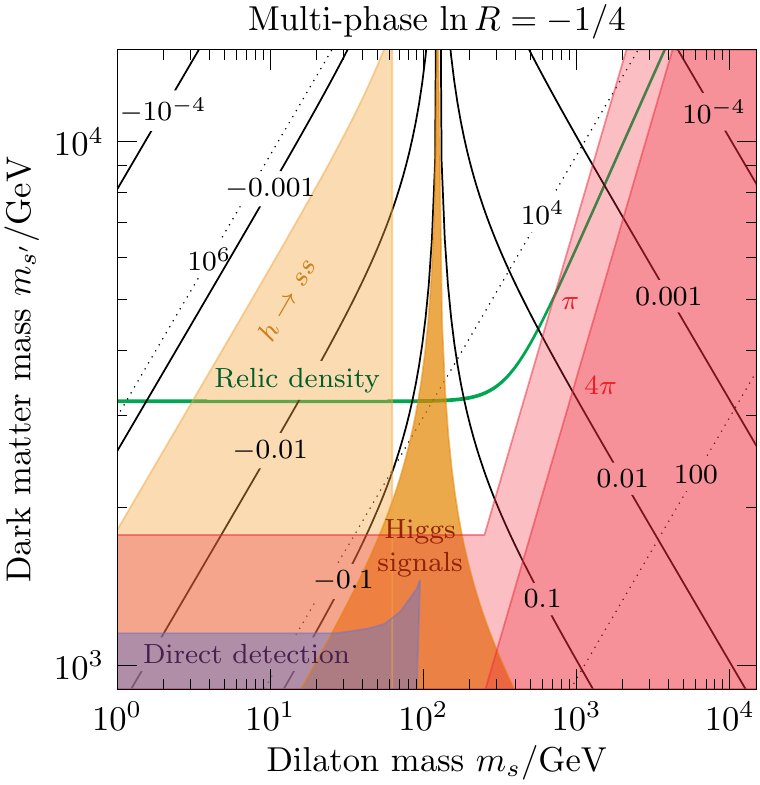}\qquad
  \includegraphics{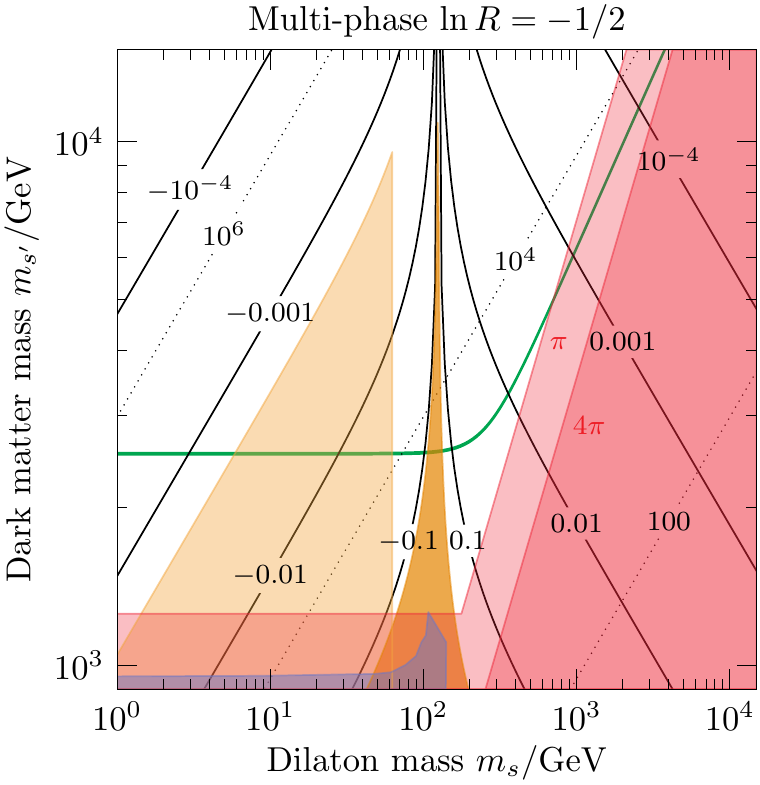}
  \\
  \vspace{2em}
   \includegraphics{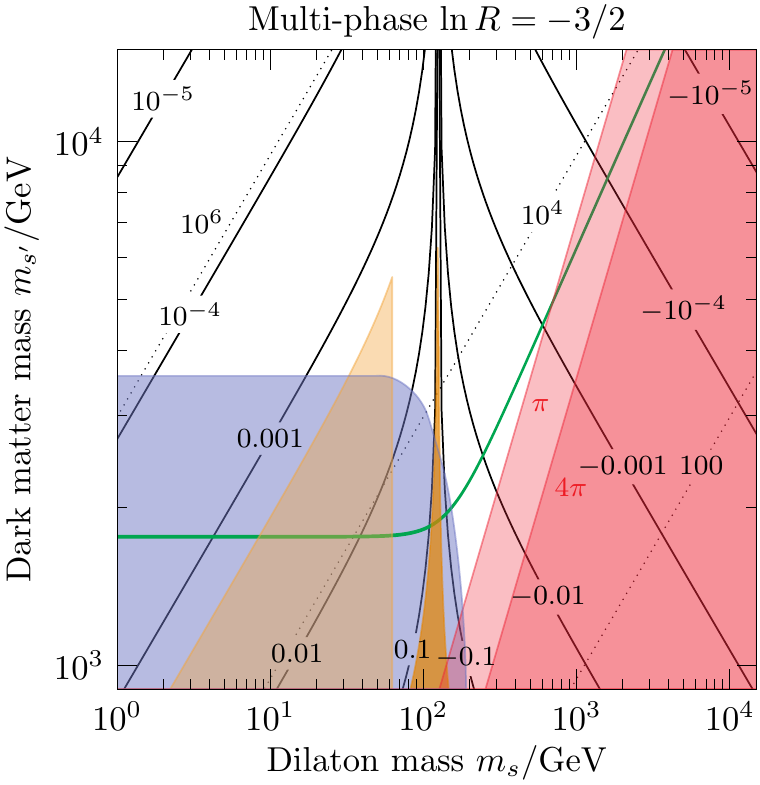}\qquad
  \includegraphics{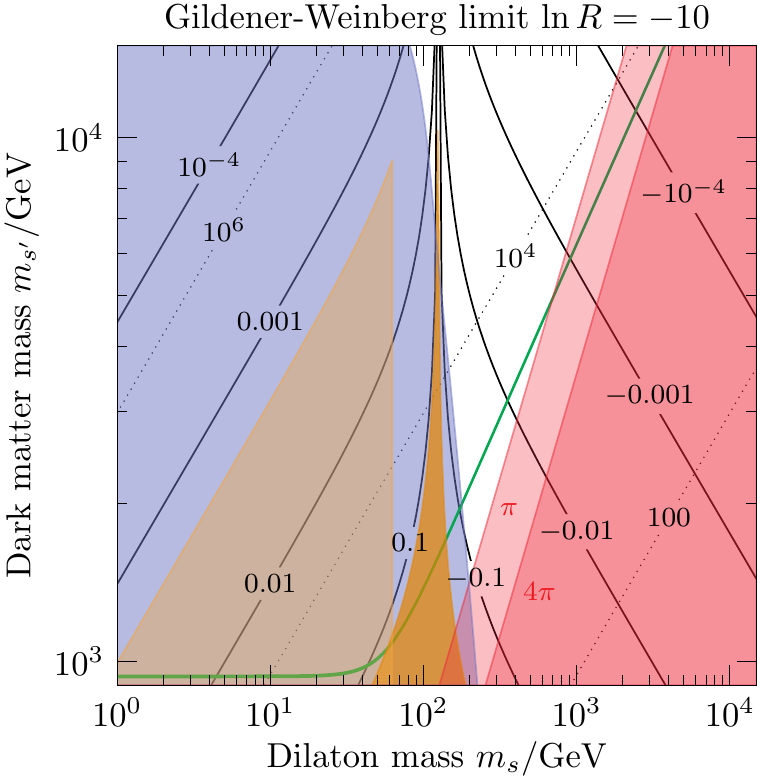}
\caption{In the (dilaton, DM) mass plane $(m_{s}, m_{s'})$ for 
$\ln R = -1/4$ (top left panel, multi-phase criticality regime), 
$\ln R = -1/2$ (top right panel, multi-phase criticality regime), 
$\ln R = -3/2$ (bottom left panel, multi-phase criticality regime), and 
$\ln R = -10$ (bottom right panel, Gildener-Weinberg limit), we show
the region where the DM cosmological abundance is reproduced via freeze-out (in green, $3\sigma$ limits). The shaded areas are excluded by
perturbativity bounds on $\lambda_{HS'}$ and $\lambda_{SS'}$ (darker and lighter red regions use a $4 \pi$ and a $\pi$ limit, respectively);
by measurements of the Higgs boson couplings (orange); by limits on the  $h \to s s$ decays contributing to the Higgs boson invisible width (beige); by direct detection experiments (purple).
In each panel, we also show the contours for the Higgs/dilaton mixing angle $\sin \theta$ (solid lines) and the dilaton VEV (dotted lines, in GeV).
\label{fig:angle}}
\end{center}
\end{figure*}

\begin{figure}[tb]
\begin{center}
  \includegraphics{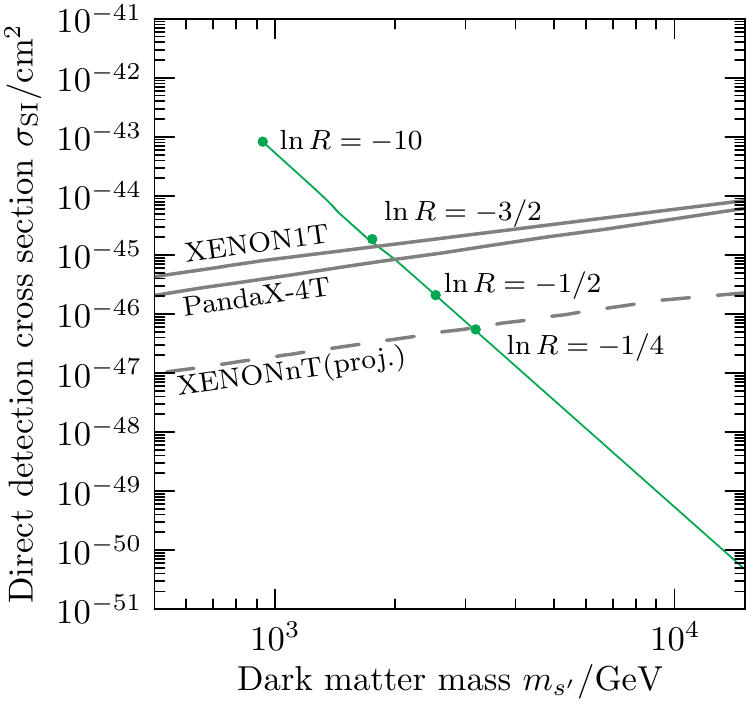}
\caption{The dark matter direct detection cross section (green), given by Eq.~\eqref{eq:dd}, together with the XENON1T(2018), PandaX-4T(2021) and the projected XENONnT sensitivity curves. The green dots correspond to cross section obtained with the minimal value of DM mass allowed by the indicated value of $\ln R$ and by the observed relic abundance.}
\label{fig:direct:detection}
\end{center}
\end{figure}


\section{Phenomenology}\label{sec:dm}
We discuss here the DM phenomenology predicted by the model and its connection with dynamical symmetry breaking.

\subsection{Cosmological abundance}
To begin with, we compute the DM mass $m_{s'}$ yielding the observed relic abundance 
$\Omega_{\rm DM} h^{2} = 0.120 \pm 0.001$~\cite{Planck:2018vyg} via thermal freeze-out. We anticipate that this requirement selects the heavy-DM limit $m_{s'}\gg m_s, m_h$, where the non-relativistic DM annihilations cross section is simply given by
\be 
\sigma_{\rm ann}  v_{\rm rel} 
\approx\frac{ \lambda_{SS'}^2+4\lambda_{HS'}^2}{64\pi m_{s'}^2}
\approx4\pi^3  \frac{m_s^4+4m_h^4/\ln^2 R}{m_{s'}^6}.
\label{eq:sigmavsimple}\ee
In fact, the annihilation cross section can be straightforwardly computed in the unbroken limit $v=w=0$, where only the Feynman diagrams using quartic interactions contribute.\footnote{Since symmetry breaking in this model
only affects energies below the DM mass, it should not affect higher-energy scatterings. We have verified that, indeed, the same cross section arises
in the broken phase in a non-trivial way: extra diagrams with cubic interactions and $t,u$-channel $s'$ mediator
contribute to the total amplitude, flipping its sign.}
There is a factor of 4 between the two annihilation modes because the Higgs multiplet has four components, while $S$ has one. DM annihilation cross sections for generic values of the DM mass $m_{s'}$ are listed in~\ref{sec:annihilation}, where we account for all the symmetry-breaking effects and the different final states.\footnote{A full numerical study is done using
the micrOMEGAs code~\cite{Belanger:2018ccd} with model files generated by the FeynRules package~\cite{Alloul:2013bka,Christensen:2008py}.}

\smallskip

As usual, the cosmological DM abundance is reproduced via freeze-out 
if 
\be \sigma_{\rm ann}  v_{\rm rel}  \approx \frac{1}{M^2}\quad\text{with}\quad
M\approx 23~{\rm TeV}.\ee
Therefore, there are two regimes depending on whether $\lambda_{SS'}$  
or $\lambda_{HS'}$ dominates in Eq.~\eqref{eq:sigmavsimple}.
Notice that according to Eqs.~\eqref{eq:DM:dilaton:portal} and~\eqref{eq:DM:Higgs:portal}, it is $\lambda_{SS'}>\lambda_{HS'}$ if the dilaton is heavier than the Higgs, 
$m_s> m_h/\sqrt{-\ln R}$. Hence, the two regimes are:
\begin{itemize}
\item If $\lambda_{HS'}$ dominates (for $m_s \ll m_h$), the DM cosmological abundance sets the DM mass to
\be m_{s'}= \sqrt{\pi} (2 m_h)^{2/3} M^{1/3}/(-\ln R)^{1/3} \label{eq:mscte}\ee
independently of $m_s$.
Then the coupling $\lambda_{HS'}$, relevant for direct detection, is forced to a relatively small value
\be \lambda_{HS'} \simeq 4\pi  (2m_h/M)^{2/3}/(-\ln R)^{1/3}
\ee
and $\lambda_{SS'}$ is a fortiori smaller.
In this regime DM behaves as in the singlet scalar model~\cite{Abada:2011qb,Abada:2012hf,Abada:2013pca}.

\item If $\lambda_{SS'}$ dominates (for $m_s \gg m_h$, i.e.\ the right side of each panel in Fig.~\ref{fig:angle}), then
DM has a larger mass that grows with $m_s$,
\be m_{s'} = \sqrt{\pi} m_s^{2/3} (2M)^{1/3},  \label{eq:msgrow}\ee
and thereby the coupling $ \lambda_{HS'}$ that controls direct detection diminishes,
\be \lambda_{HS'} \simeq  \frac{1024^{1/3}\pi m_h^2}{M^{2/3} m_s^{4/3}(-\ln R)},
\label{eq:mscte:hi}
\ee
while $\lambda_{SS'} \simeq 1024^{1/3}\pi (m_s/M)^{2/3}$ increases,
eventually conflicting with perturbativity bounds.
\end{itemize}

These two limits explain the numerical results shown in 
Fig.~\ref{fig:angle}, where the four panels correspond to growing values of
$-\ln R = 1/4,\, 1/2, \, 3/2, \,10$, chosen to show how the
Gildener-Weinberg approximation is progressively approached.
The green bands denote the DM masses indicated by thermal freeze-out:
the left side of each panel ($m_s < m_h$) shows a constant DM mass $m_s$ in agreement with Eq.~\eqref{eq:mscte},
and the right side of each panel ($m_s>m_h)$ shows a growing DM mass in agreement with Eq.~\eqref{eq:msgrow}. 
No other disjoint solutions --- e.g. at resonances $m_{s'}=m_{s,h}/2$ ---  match the observed DM relic density. Below the green curves, $s'$ contributes only to a fraction of the observed DM abundance. Conversely, the region above the green curve, corresponding to a DM over-density, is excluded. The areas shaded in red signal where a coupling exceeds the non-perturbative limit ($4 \pi$ for darker red, $\pi$ for lighter red). In the slanted region on the right of each plot this happens because of $\lambda_{SS'}$; in the horizontal region at the bottom, instead, $\lambda_{HS'}$ violates the bound. The orange area shows the bound from a fit of the Higgs couplings and the beige is excluded by searches targeting the $h \to s s$ decay~\cite{Robens:2016xkb}. The contour lines show the Higgs-dilaton mixing angle $\sin \theta$ (solid lines) and the dilaton vacuum expectation value $w/\text{GeV}$ (dotted lines). In the light purple area, the scaled direct detection cross section $(\Omega_{s'}/\Omega_{\rm DM}) \sigma_{\rm SI}$ exceeds the PandaX-4T(2021) limit~\cite{PandaX-4T:2021bab}. 

\smallskip

Eq.~\eqref{eq:sigmavsimple} also gives the cross section for indirect DM detection signals:
DM annihilates with thermal cross section either into $h,W,Z$ (if $\lambda_{HS'}$ dominates)
or into the dilaton that next decays into SM particles (if $\lambda_{SS'}$ dominates),
giving the corresponding spectra of long-lived light SM particles \cite{Cirelli:2010xx}.

\subsection{Direct detection}\label{sec:direct}
DM scatters on nuclei via the Higgs boson, which interacts with nucleons $N$ through the effective coupling
\be
 \frac{f_N m_N}{v} h\bar N N,
\ee
where $m_N=0.946$ GeV is the nucleon mass and $f_N\approx 0.3$ is a form factor~\cite{Alarcon:2011zs, Alarcon:2012nr, Cline:2013gha}. 
After taking into account the mass mixing of $h$ with the dilaton $s$,
both mass eigenstates $h_{1,2}$ contribute to the spin-independent (SI) direct detection cross section. At tree level, in the low energy limit, we therefore have
\begin{align}
\sigma_{\rm SI} = \frac{f_N^2 m_N^2 \mu^2}{\pi m_{s'}^2 v^2} 
\left[\frac{\lambda_{h_1 s' s'}}{m_1^2}\cos\theta +\frac{\lambda_{h_2 s' s'}}{m_2^2}\sin\theta\right]^2,
\end{align}
where 
$\mu=m_{s'}m_N/(m_{s'}+m_N) \simeq m_N$ is the reduced DM/nucleon mass,
$\theta$ is the mixing angle that defines the two eigenstates, 
and $\lambda_{h_i s' s'}$ are the corresponding couplings listed in \ref{sec:annihilation}.

By retaining only the linear contribution in the small mixing angle, the direct-detection cross section simplifies to
\begin{align}
\sigma_{\rm SI}\simeq \frac{f_N^2 m_N^4 }{4\pi m_{s'}^2}\Bigg[
\frac{\lambda_{HS'}}{m_h^2}+\frac{\lambda_{SS'}}{m_s^2}\frac{1+\ln R}{\ln R}\Bigg]^2, \end{align}
matching the mass-insertion approximation. Since all scalars acquire mass through their interactions, the model relates the couplings to the particle masses. Hence,  by inserting the expressions for these quantities respectively given in Eqs.~\eqref{eq:DM:dilaton:portal}, \eqref{eq:DM:Higgs:portal} and Eqs.~\eqref{eq:m:sq:h}, \eqref{eq:m:sq:s},
the $h$-mediated and $s$-mediated contributions simplify in a total that only depends on the DM mass $m_{s'}$:  
\begin{align}
\sigma_{\rm SI}\simeq
\frac{64 \pi^3 f_N^2 m_N^4}{m_{s'}^6}.
\label{eq:dd}
\end{align}
The result is shown by the green line in Fig.~\ref{fig:direct:detection}.
Thereby, the plot shows the XENON1T(2018)~\cite{XENON:2018voc} and PandaX-4T(2021)~\cite{PandaX-4T:2021bab} bounds, as well as the projected XENONnT~\cite{XENON:2020kmp} sensitivity. As we can see, current experimental bounds are satisfied for DM heavier than $m_{s'}\gtrsim2\,{\rm TeV}$. In particular, this happens in the heavy dilaton regime $m_{s} \gg m_{h}$, where the DM relic density is controlled by the $\lambda_{SS'}$ coupling, while $\lambda_{HS'}$ is given by Eq. \eqref{eq:mscte:hi}. The direct detection constraint thus allows a range of $m_{s'}$ for which the  cosmological abundance can be reproduced by
thermal freeze-out. As discussed previously, for each value of $-\ln R$ this implies a minimal value of $m_{s'}$. The dots in Fig.~\ref{fig:direct:detection} mark these values. Approaching the Gildener-Weinberg limit of large $|\ln R|$, DM increasingly lightens. This conflicts with the direct detection results, as signaled by the larger extent of purple regions in the last two panels of Fig.~\ref{fig:angle}.

\begin{figure}[tb]
  \begin{center}
    \includegraphics{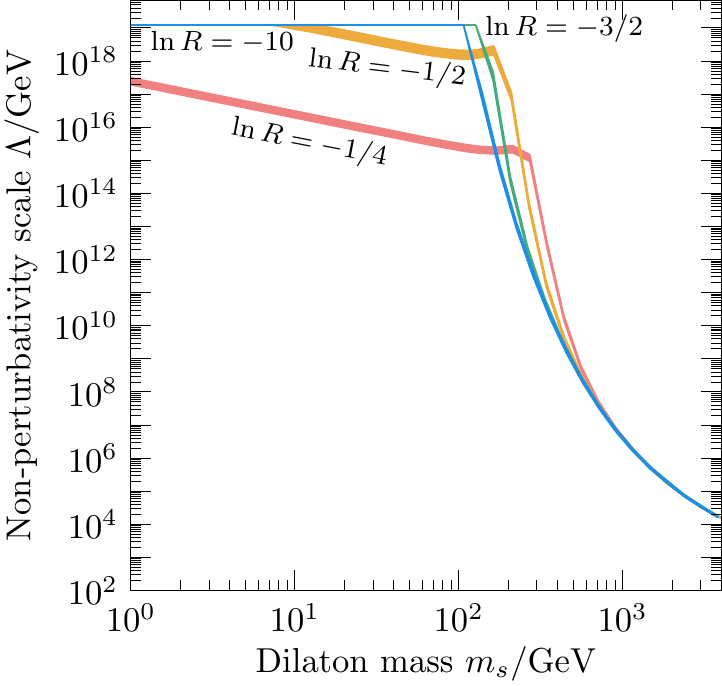}
  \caption{The scale up to which the model remains perturbative in the parameter space that satisfies the Planck DM relic density at a $3 \sigma$ level.}
  \label{fig:pert:scale}
  \end{center}
  \end{figure}

\subsection{Perturbativity scale}\label{sec:pert}
In order to investigate the perturbativity scale of the model, we use the PyR@te 3 package~\cite{Sartore:2020gou} to derive the full expressions of the quartic coupling $\beta$-functions. The perturbativity scale is estimated as the lowest of the RG scale values at which the couplings cross the $4 \pi$ threshold during their running. The obtained perturbativity scales (capped at the Planck scale) are shown in Fig.~\ref{fig:pert:scale} for the indicated values of $\ln R$ and dilaton mass $m_{s}$.  In our analysis we set $m_{s'}$ so as to satisfy the relic density constraint and $\lambda_{S'}=0$ at the electroweak scale. As we can see, the model remains valid up to the Planck scale for $-\ln R = 1/2, 3/2, 10$ provided that the dilaton is lighter than the Higgs boson.

\section{Conclusions}\label{sec:concl}

We have studied a classically scale-invariant model in which dynamical symmetry breaking is driven by the interactions with dark matter. The scalar sector of the theory contains, besides the standard model Higgs boson, two gauge singlets: the dilaton and dark matter. We find a tight connection between dark matter and Higgs boson phenomenology.

In our analysis we have paid particular attention to novel solutions arising in the multi-phase criticality limit. In this regime some radiative corrections sourced by dark matter, which are neglected in the usual Gildener-Weinberg approach, are crucial to the dynamics of symmetry breaking. Both the dilaton and the Higgs boson acquire loop-suppressed masses that fall well below the scale of symmetry breaking. 
The model can be parameterised by three free parameters: the mass of the dilaton $m_{s}$, the dark matter mass $m_{s'}$ and a parameter $R$ that measures the deviation from the usual Gildener-Weinberg limit (obtained for large $\abs{\ln R}$).

The dark matter direct detection cross section, Eq.~\eqref{eq:dd}, takes an especially simple form because the Higgs boson and the dilaton are coupled to dark matter via the same interactions that generate their masses through dynamical symmetry breaking.

The model satisfies the current bounds pertaining to dark matter and Higgs boson phenomenology and could be tested in future direct detection experiments such as the XENONnT.

\subsection*{Acknowledgement}\footnotesize
This work was supported by European Regional Development Fund through the CoE program grant TK133, 
by the Mobilitas Pluss grants MOBTT5, MOBTT86, by the Estonian Research Council grants PRG434, PRG803 and PRG356,
and by Italian MIUR under PRIN 2017FMJFMW.

\appendix

\section{Full parametrization of the quartic couplings}
\label{sec:exact}
The entries of the scalar mass matrix are 
\begin{align}
  m^2_{hh} & = - \frac{\beta_{\lambda_{HS}}}{(4 \pi)^{2}} w^2 \ln R,
  \label{eq:m:hh} 
  \\
  m^2_{ss} & =  2 \frac{\beta_{\lambda_S}}{(4 \pi)^{2}} w^2  - \frac{\beta_{\lambda_{HS}}^{2} \ln R \, (3 + \ln R)}{4 \lambda_{H} [(4 \pi)^{2}]^{2}} w^2,
  \label{eq:m:ss}
  \\
  m^2_{hs} & = w^2 \sqrt{-\left[\frac{\beta_{\lambda_{HS}} }{(4 \pi)^{2}} \right]^{3} \ln R} \; \frac{ (1 + \ln R)}{ \sqrt{2 \lambda_{H}}},
  \label{eq:m:h:s}
\end{align}
whereas the full expression of the $s'$ mass is 
\begin{equation}
  m^{2}_{s'} = \frac{1}{2} \lambda_{SS'} w^{2} - \frac{1}{2} \frac{\lambda_{HS'} \beta_{\lambda_{HS'}} \ln R}{2 (4 \pi)^{2} \lambda_{H}} w^{2}.
  \label{eq:m:spsp}
\end{equation}
By substituting the $\beta$-functions of Eq.~\eqref{eq:beta:at:s0} in the Eqs.~\eqref{eq:m:hh}, \eqref{eq:m:ss} and \eqref{eq:m:spsp}, we obtain in the limit of small mixing angle that
\begin{align}
  \lambda_{SS'} &\approx 64 \pi^{2} m_{s}^{2} / {\Big[} m_{s'}^{2} \ln R (5  + \ln R)
  \notag
  \\
  & + (1 + \ln R) \sqrt{\ln R (m_{s'}^{4} \ln R + 32 \pi^{2} v^{2} m_{h}^{2})}
  {\Big ]},
  \label{eq:DM:dilaton:portal:compl}
\\
  \lambda_{HS'} &\approx \frac{m_{s'}^{2} \ln R + \sqrt{\ln R (m_{s'}^{4} \ln R + 32 \pi^{2} v^{2} m_{h}^{2})}}{v^{2} \ln R}.
  \label{eq:DM:Higgs:portal:compl}
\end{align}
These solutions are real if the DM mass $m_{s'}$ is larger than
\begin{equation}
  m_{s'} \ge 2 \sqrt{\pi m_{h} v} \left(\frac{2}{-\ln R}\right)^{{1}/{4}},
  \label{eq:msp:lower:bound}
\end{equation}
in which case the interactions with DM affect dynamical symmetry breaking through the RG running of the involved couplings. The bound holds in the parameter space considered in our analysis and asymptotically vanishes in the Gildener-Weinberg limit $\ln R \to -\infty$.

\section{Renormalization group equations}
\label{sec:RGEs}
The full one-loop expressions of the $\beta$-functions are  
\begin{align}
  \beta_{\lambda_H} &= \beta_{\lambda_H}^{\rm SM} + \frac{1}{4} (\lambda_{HS}^2 +  \lambda_{HS'}^2),
  \\
  \beta_{\lambda_S} &= 9 \lambda_{S}^2 +  \frac14 \lambda_{SS'}^2  +  \lambda_{HS}^2,
  \\
  \beta_{\lambda_{S'}} &= 9\lambda_{S'}^2  + \frac14 \lambda_{SS'}^2 +  \lambda_{HS'}^2,
  \\
  \beta_{\lambda_{HS}} &= \lambda_{HS} (\tilde Z_h  + 3\lambda_S+2\lambda_{HS})+\frac12 \lambda_{SS'} \lambda_{HS'},
  \\
 \beta_{\lambda_{HS'}} &= \lambda_{HS'}(\tilde Z_h  + 3\lambda_{S'}+2\lambda_{HS'})  + \frac12 \lambda_{SS'} \lambda_{HS},
  \\
 \beta_{\lambda_{SS'}} &=  \lambda_{SS'} (2 \lambda_{SS'} + 3 \lambda_{S}+
      3 \lambda_{S'}) + 2 \lambda_{HS} \lambda_{HS'},
\end{align}
where $\tilde Z_h = Z_h + 6 \lambda_{H}$, and
\begin{equation}
Z_h = 3 y_{t}^2- \frac{9}{4} g_2^2  -\frac 34  g_Y^2
\end{equation}
is the one-loop wave-function renormalisation for the Higgs boson.

\section{DM annihilation cross sections}
\label{sec:annihilation}

In this Appendix we show the full expression for the DM annihilation cross sections computed at tree-level. The scalar couplings in the eigenstate basis $h_1$ (mostly Higgs) and $h_2$ (mostly dilaton) are
\begin{align}
&\lambda_{h_1 s' s'} = \frac{1}{2}\Big(-v\lambda_{HS'}\cos\theta+ w\lambda_{SS'}\sin\theta\Big),\\
&\lambda_{h_2 s' s'} = -\frac{1}{2}\Big(w\lambda_{SS'}\cos\theta+ v\lambda_{HS'}\sin\theta\Big),\\
& \lambda_{h_1 h_1 s' s'} = -\frac{1}{4}\Big(\lambda_{HS'}\cos^2\theta + \lambda_{SS'}\sin^2\theta\Big),\\
& \lambda_{h_2 h_2 s' s'} = -\frac{1}{4}\Big(\lambda_{SS'}\cos^2\theta + \lambda_{HS'}\sin^2\theta\Big),\\
& \lambda_{h_1 h_2 s' s'} = \frac{1}{2}\Big(-\lambda_{HS'} + \lambda_{SS'}\Big)\cos\theta\sin\theta.
\end{align}
where $\theta$ is the mixing angle. Henceforth, we will use the abbreviations $s_\theta\equiv \sin \theta$, and $c_\theta\equiv \cos \theta$ and the definitions
\begin{align}
& k_i=\frac{m_i^2}{s},\quad \beta_i=\sqrt{1-\frac{4m_i^2}{s}},\\
& \beta_{ij}=\sqrt{1-\frac{2(m_i^2+m_j^2)}{s}+\frac{(m_i^2-m_j^2)^2}{s^2}},
\end{align}
yielding $v_{\rm rel}=2\beta_{s'}$.

The annihilation cross sections into scalar final states are
\begin{align}
&\sigma_{s' s' \to h_i h_i}
=\frac{1}{4\pi}\frac{1}{v_{\rm rel}}\frac{\beta_{i}}{s}\Bigg\{\nonumber\\
& \alpha^2_i
+ \alpha_i \frac{8\lambda_{h_i s' s'}^2 }{s\beta_{s'}\beta_{i}}\log\left(\frac{1-2k_{i}+\beta_{s'}\beta_{i}}{1-2k_{i}-\beta_{s'}\beta_{i}}\right)
\notag\\
& + \frac{16\lambda_{h_i s' s'}^4}{s^2}\Bigg[-\frac{2}{\beta_{s'}^2\beta^2_{i}-(1-2k_{i})^2}\nonumber\\
&+\frac{1}{\beta_{s'}\beta_{i}(1-2k_{i})}\log\left(\frac{1-2k_{i}+\beta_{s'}\beta_{i}}{1-2k_{i}-\beta_{s'}\beta_{i}}\right)\Bigg]\Bigg\} ,
\end{align}
and
\begin{align}
& \sigma_{s' s'\to h_1 h_2}
=\frac{\beta_{12}}{4\pi v_{\rm rel}s}\Bigg\{\nonumber\\
&\alpha_{12}^2+\alpha_{12}\frac{16\lambda_{h_1 s' s'}\lambda_{h_2 s' s'}}{s\beta_{s'}\beta_{12}}\log\left(\frac{1-k_{1}-k_{2}+\beta_{s'}\beta_{12}}{1-k_{1}-k_{2}-\beta_{s'}\beta_{12}}\right)
\notag\\
&+\frac{32\lambda_{h_1 s' s'}^2\lambda_{h_2 s' s'}^2}{s^2}\left[-\frac{2}{\beta_{s'}^2\beta_{12}^2-(1-k_1-k_2)^2}
\right.
\notag
\\
&+ \left. \frac{1}{\beta_{s'}\beta_{12}(1-k_1-k_2)}\log\left(\frac{1-k_{1}-k_{2}+\beta_{s'}\beta_{12}}{1-k_{1}-k_{2}-\beta_{s'}\beta_{12}}\right)\right]\Bigg\},
\end{align}
where
\begin{align}
&\alpha_1=2\lambda_{h_1 h_1 s' s'}-\frac{6\lambda_{h_1 s' s'}\lambda_{h_1 h_1 h_1}}{s-m_1^2}
-\frac{2\lambda_{h_2 s' s'}\lambda_{h_2 h_1 h_1}}{s-m_2^2},\\
&\alpha_2=2\lambda_{h_2 h_2 s' s'}-\frac{2\lambda_{h_1 s' s'}\lambda_{h_1 h_2 h_2}}{s-m_1^2}
-\frac{6\lambda_{h_2 s' s'}\lambda_{h_2 h_2 h_2}}{s-m_2^2},\\
&\alpha_{12}=\lambda_{h_1 h_2 s' s'}-\frac{2\lambda_{h_1 s' s'}\lambda_{h_2 h_1 h_1}}{s-m_1^2}
-\frac{2\lambda_{h_2 s' s'}\lambda_{h_1 h_2 h_2}}{s-m_2^2}.
\end{align}
The annihilation cross sections into massive vector bosons (in unitary gauge) are
\begin{align}
&\sigma_{s's'\to W^+ W^-}=\frac{\beta_W}{2\pi v_{\rm rel} s v^2}\left[s^2-4m_W^2 s+12 m_W^4\right]\nonumber\\
&\quad\quad\times\left[\frac{c_\theta \lambda_{h_1 s' s'}}{s-m_1^2}+\frac{s_\theta \lambda_{h_2 s' s'}}{s-m_2^2}\right]^2,\\
&\sigma_{s' s' \to ZZ}=\frac{\beta_Z}{4\pi v_{\rm rel} s v^2}\left[s^2-4m_Z^2 s+12 m_Z^4\right]\nonumber\\
&\quad\quad\times\left[\frac{c_\theta \lambda_{h_1 s' s'}}{s-m_1^2}+\frac{s_\theta \lambda_{h_2 s' s'}}{s-m_2^2}\right]^2,
\end{align}
while the subdominant fermion channels give
\be
\sigma_{s's'\to \bar{f}f}=\frac{N_c \beta_{f}^3 m_f^2}{\pi v_{\rm rel} v^2} \left[\frac{c_\theta \lambda_{h_1 s' s'}}{s-m_1^2}+\frac{s_\theta \lambda_{h_2 s' s'}}{s-m_2^2}\right]^2,
\ee
where $N_c=1$ for leptons and $3$ for quarks.

\bigskip\small\normalsize

\bibliographystyle{elsarticle-num}
\bibliography{artCW}

\end{document}